\documentclass[aps,prl,twocolumn,nofootinbib,preprintnumbers,superscriptaddress]{revtex4}

\def\mop#1{\mathop{\rm #1}\nolimits}
\def\tr{\mop{tr}}

\begin{document}

\preprint{COLO-HEP-569, PUPT-2400}

\title{Fermi surfaces in maximal gauged supergravity}

\author{Oliver DeWolfe}
\affiliation{Department of Physics, 390 UCB, University of Colorado, Boulder, CO 80309, USA}
\author{Steven S. Gubser}
\affiliation{Joseph Henry Laboratories, Princeton University, Princeton, NJ 08544, USA}
\author{Christopher Rosen}
\affiliation{Department of Physics, 390 UCB, University of Colorado, Boulder, CO 80309, USA}

\date{\today}

\begin{abstract}
We obtain fermion fluctuation equations around extremal charged black hole geometries in maximal gauged supergravity in four and five dimensions, and we demonstrate that their solutions display Fermi surface singularities for the dual conformal field theories at finite chemical potential.  The four-dimensional case is a massless charged fermion, while in five dimensions we find a massive charged fermion with a Pauli coupling.  In both cases, the corresponding scaling exponent is less than one half, leading to non-Fermi liquid behavior with no stable quasiparticles, although some excitations have widths more than ten times smaller than their excitation energy. In the five-dimensional case, both the Fermi momentum and the scaling exponent appear to have simple values, and a Luttinger calculation suggests that the gauginos may carry most of the charge of the black hole.
\end{abstract}

\maketitle

\section{\label{sec:intro} Introduction}

Ordinary metallic states may be described as Fermi liquids, with dynamics characterized by weakly-coupled, long-lived quasiparticle excitations around a Fermi surface.  Certain strongly correlated electron systems, however, are different: the ``strange metals" arising in high-$T_c$ cuprates \cite{Varma, Anderson} and heavy fermion systems \cite{Gegenwart}, for example, are shown by photoemission experiments to possess a Fermi surface, but the associated gapless modes are not long-lived.  Understanding such ``non-Fermi liquids" is an important theoretical challenge.

The gauge-string duality  \cite{Maldacena:1997re,Gubser:1998bc,Witten:1998qj} is a promising avenue to investigate such phenomena.
The duality relates quantum field theory without gravity ``holographically" to string theory or supergravity in higher dimensional spacetimes.  Crucially, strongly-coupled systems which are difficult to characterize are mapped onto black hole geometries which can be far more tractable calculationally;  such techniques have had  success describing the quark-gluon plasma (see {\em e.g.}~\cite{CasalderreySolana:2011us}).
Since the first work on obtaining holographic Fermi surfaces \cite{Lee:2008xf,Liu:2009dm, Cubrovic:2009ye}, black hole geometries describing non-Fermi liquids have indeed been obtained;
 an investigation of fermion behavior for general dimension, mass and charges was carried out in \cite{Faulkner:2009wj}. 
Such studies have typically followed a ``bottom-up" approach: rather than finding an explicit solution of string theory or supergravity, instead a convenient effective gravity Lagrangian is postulated and its consequences determined.

While this approach is highly valuable, it has drawbacks.  
Without an embedding in string theory, the precise nature of the dual quantum field theory is unknown, and one is left ignorant of exactly what system is supporting the Fermi surface.
Furthermore, one may worry that the results could be an artifact of unphysical parameter choices on the gravity side, which might not be present in any construction descending from string theory.
Therefore, a ``top-down" approach, where one starts from a known string or supergravity construction and investigates fermionic behavior, is naturally valuable both for understanding the system and having confidence in its validity.
 Previous top-down approaches to Fermi surfaces include \cite{Ammon:2010pg, Jensen:2011su}, which studied fermions realized on probe branes, and \cite{Gauntlett:2011mf, Belliard:2011qq, Gauntlett:2011wm}, which studied the gravitino in the gravity multiplet of minimal supergravity and found no Fermi surface singularity.

In this work, we study spin-1/2 fermions in both four-dimensional and five-dimensional maximally supersymmetric gauged supergravity, and we find Fermi surfaces in both cases.  These theories descend from M-theory on $S^7$ and type IIB string theory on $S^5$, respectively, and the dual field theories are the ones describing stacks of M2-branes and stacks of D3-branes in otherwise empty spacetime.  The D3-brane theory is ${\cal N}=4$ super-Yang-Mills theory, which is by far the best-studied example of the gauge-string duality.

In both cases, we find non-Fermi liquid behavior with excitations whose width is comparable to their energy, resulting in no well-defined quasiparticles close to the Fermi surface.  At least some excitations have a width more than ten times smaller than the energy for small energies, indicating they may be fairly well-defined as resonances.  
For the case of ${\cal N}=4$ super-Yang-Mills, we find simple values for the Fermi momenta, and an unusual $\omega \sim k_\perp^6$ dispersion relation for excitations around the Fermi surfaces. The singularity in the Green's function corresponding to the Fermi surface has a residue of order $N^2$, where $N$ is the number of colors in the gauge theory.  This suggests that the relevant fermionic field theory degrees of freedom are in the adjoint of the $SU(N)$ gauge group---not color-singlet bound states as has been proposed for certain bottom-up constructions.  A Luttinger count of the charge density indicates most charge is carried by the adjoint gaugino fields. Our entire field theory analysis is only possible because of the knowledge of the dual theory stemming from the top-down construction.

We proceed through three steps for each case.  First, we recall how one identifies a subset of the fields of the gauged supergravity into which one can embed an extremal Reissner-Nordstr\"om anti-de Sitter (RNAdS) black hole; this is nontrivial since turning on a generic gauge field will source unwanted scalars as well.  Second, we isolate particular charged fermions and derive their linearized fluctuation equations.  Third, we solve these equations in the extremal RNAdS background with suitable infalling boundary conditions, identify a Fermi surface by finding the momentum at which the ``source" term vanishes, and study the excitations near this surface.

\section{\label{sec:embed} Fermi surfaces from four-dimensional supergravity}

The extremal RNAdS solution in $D = d+1$ dimensions is
\begin{eqnarray}\label{RNAdS}
ds^2 &=& {r^2 \over L^2} (f dt^2 - d\vec{x}^2) - {L^2 \over r^2} {dr^2 \over f} \,, \\
a_\mu dx^\mu &=& \mu \left( 1 - \left( r_0 \over r \right)^{d-2} \right) dt \,,
\nonumber\end{eqnarray}

with the horizon function
\begin{equation}\label{HorizonFunction}
f = 1 + {d \over d-2} \left( r_0 \over r \right)^{2d-2} - {2 (d-1) \over  (d-2)} \left( r_0 \over r \right)^d 
 \,, 
\end{equation}
where $r_0$ is the horizon radius $f(r_0) \equiv 0$.

In order to embed the $D=4$ charged black hole (\ref{RNAdS}) into four-dimensional gauged $\mathcal{N} = 8$ supergravity \cite{deWit:1982ig}, we start with the $SO(8)$ gauge fields $A_{\mu[ij]}$, where $i$ and $j$ run from $1$ to $8$, and set all of them to $0$ except\footnote{We note that choosing only $A_{\mu12}$ non-zero is related to choosing all four Cartan gauge fields  non-zero and equal $A_{\mu12} = A_{\mu34} = A_{\mu56} = A_{\mu78}$ by an $SO(8)$ triality transformation.}
\begin{equation}\label{4DGauge}
a_\mu \equiv A_{\mu12} = -A_{\mu21}\,.
\end{equation}
The supergravity scalar fields will remain in the trivial configuration throughout, corresponding to an undeformed $S^7$ in the lift to eleven-dimensional supergravity.  The bosonic lagrangian is
\begin{equation}\label{4DAction}
e^{-1} {\cal L} = -{1 \over 2} R - {1 \over 4} f_{\mu\nu} f^{\mu\nu} + 6 g^2 \,,
\end{equation}
where $f_{\mu\nu} \equiv \partial_\mu a_\nu - \partial_\nu a_\mu$; note that we use the mostly-minus metric conventions of \cite{deWit:1982ig}.  In order for (\ref{RNAdS}) to satisfy the equations of motion from (\ref{4DAction}), one must require
\begin{equation}
g={1\over \sqrt{2} L} \,, \qquad\qquad
\mu = {\sqrt{6} r_0 \over L} \,,
\end{equation}
corresponding in the notation of  \cite{Faulkner:2009wj} to $g_F = \sqrt{2}L$.

We now consider the spin-1/2 fermions of four-dimensional ${\cal N}=8$ gauged supergravity: there are 56 Majorana spinors $\chi_{ijk}= \chi_{[ijk]}$ where $i,j,k = 1 \ldots 8$.  With scalars remaining in their trivial, symmetry preserving configuration, the covariant derivative is
\begin{equation}
\label{Dchi}
D_\mu \chi_{ijk} = \nabla_\mu \chi_{ijk} + 3 g A_{\mu\;\;\;[i}^{\;\;m}\, \chi_{jk]m} \,.
\end{equation}
The 56 fermions divide into three sectors: 6 containing both the indices $1$ and $2$, 20 containing neither, and 30 containing just one of them.  It is easy to see that only the last sector is charged under $a_\mu$.

The terms in the ${\cal N}=8$ Lagrangian that contribute to the quadratic fermion action are 
\begin{eqnarray}
e^{-1}{\cal L}_{1/2} &=& - {1 \over 12}  \bar\chi^{ijk}(\gamma^\mu D_\mu- \overleftarrow{D}_\mu \gamma^\mu)  \chi_{ijk}  \\ && - {1 \over 2} \left(F^+_{\mu\nu ij} S^{ij, kl} O^{+\mu\nu kl} + {\rm h.c.} \right)\,,
\nonumber
\end{eqnarray}
where  $S^{ij,kl} = \delta^{[i}_j \delta^{k]}_l$ when the scalars are trivial, and $O^{+\mu\nu ij}$ is bilinear in the $\chi_{ijk}$ and the gravitini $\psi_{\rho i}$:
\begin{eqnarray}
O^{+\mu\nu ij} &\equiv& - {\sqrt{2} \over 144} \epsilon^{ijklmnpq}\bar\chi_{klm} \sigma^{\mu\nu} \chi_{npq}  \\ && - {1\over 2} \bar\psi_{\rho k}\sigma^{\mu\nu} \gamma^\rho \chi^{ijk} 
+ ( \psi_\rho^2 \ {\rm term}) \,.
\nonumber
\end{eqnarray}
Note there are no elementary mass terms for the spin-$1/2$ fermions.  It is easy to see that the Pauli-type terms in the presence of the gauge field couple the first sector of neutral fermions to the gravitini and generate an effective mass term for the second neutral sector, but do not affect the charged spin-1/2 fields.  Thus the 15 complex fermions of the form $\chi \sim \chi_{1jk} + i \chi_{2jk}$
with $j,k = 3, \ldots 8$ all satisfy the simple Dirac equation,
\begin{equation}\label{eq:DiracEQ}
\gamma^{\mu}\left(\nabla_\mu-i q a_\mu\right)\chi = 0 \,,
\end{equation}
with $q = g = 1/(\sqrt{2} L)$; their conjugates have charge $-q$.

We now consider the solutions to this equation.  The near-boundary behavior is controlled by the (vanishing) mass and is (see for example \cite{Iqbal:2009fd}),
\begin{eqnarray}
\chi_+ &\sim& A r^{-3/2} + B r^{-5/2} \,,\\
\chi_- &\sim& C r^{-5/2} + D r^{-3/2} \,,
\end{eqnarray}
where $\chi_\pm$ are eigenvectors of $\gamma^r$; $A$ is the source term and $D$ is the response, while $B$ and $C$ are determined by the other two.  The near-horizon behavior is
\begin{equation} \label{Horizon}
\chi \sim   (r - r_0)^{-1/2 \pm \nu_k} \,,
\end{equation}
where $\nu_k$ is the scaling exponent \cite{Faulkner:2009wj},
\begin{equation}
\nu_k=
 \frac{1}{2\sqrt{3}}\sqrt{6\left(\frac{k}{\mu\,q}\right)^2-1} \,,
\nonumber\end{equation}
which controls the IR behavior of the Green's function.  An ``oscillatory region" with imaginary $\nu_k$ thus exists for $|k| < \mu q /\sqrt{6}$.  

To find a Fermi surface we set $\omega = 0$ and impose regular boundary conditions at the horizon by choosing the plus in (\ref{Horizon}), and search for values of the spatial momentum $k$ outside the oscillatory region where the source term $A$ vanishes.  This case has in fact been treated by \cite{Faulkner:2009wj}, with $m=0$ and $q g_F = 1$; see also \cite{Hartman:2010fk} for an analytic treatment.  Setting $L = r_0 = 1$, we verify that there is a Fermi surface at $k_F \approx 0.9185$, or ${k_F \over \mu q} \approx 0.5305$, at which value the scaling exponent becomes
\begin{equation}
\nu_{k_F} \approx 0.2393 \,.
\end{equation}
The conjugate fermions with charge $-q$ see the Fermi surface at $k_F \approx -0.9185$, with the same $\nu_{k_F}$.

Because $\nu_{k_F} < 1/2$, the conformal dimension $\delta_{k_F} = {1 \over 2} + \nu_{k_F}$ in the auxiliary $AdS_2$ theory controlling the IR dynamics is less than $1$, indicating a relevant operator and non-Fermi liquid behavior.  Thus there is no Fermi velocity; in the notation of \cite{Faulkner:2009wj}, the retarded Green's function near the Fermi surface (that is, for small $\omega$ as well as small $k_\perp \equiv k-k_F$) takes the form
\begin{equation}\label{GreensFunction}
 G_R = {h_1 \over k_\perp - h_2 e^{i\gamma_F} \omega^{2\nu_{k_F}}} \,,
\end{equation}
where the real positive constants $h_1$ and $h_2$ encode ultraviolet data but the phase $e^{i\gamma_F}$ is entirely determined by the behavior of a near-horizon $AdS_2 \times R^2$ Green's function.  Since $h_2 e^{i\gamma_F}$ provides both the leading real and imaginary parts in the dispersion relation, the width of would-be quasiparticles is generically of the same order as the excitation energy, making them not well-defined. 

It is interesting to note, however, that for some of our excitations the widths are relatively small.  For negative $q$, we find $\gamma_F \approx 0.163$.
Expressing 
the fermionic quasinormal frequency lying closest to the Fermi surface as $\omega_{\rm QNM} = \omega_* - i\Gamma$, we find for particles ($k_\perp > 0$):
\begin{equation}
 {\Gamma \over \omega_*} = \tan {\gamma_F\over 2 \nu_{k_F}} \approx {1 \over 2.8}\,, \quad \quad k_\perp > 0 \,,\label{4DBigWidth}
\end{equation}
while for holes ($k_\perp < 0$),
\begin{equation}
 {\Gamma \over \omega_*} = \tan {\gamma_F- \pi \over 2 \nu_{k_F}}  \approx {1 \over 16.8}\,,\quad \quad k_\perp < 0 \,, \label{4DSmallWidth}
\end{equation}
manifesting a particle-hole asymmetry characteristic of $\nu_{k_F} < 1/2$ models \cite{Faulkner:2009wj}.
The positive $q$ fermion sees the Fermi surface with particles and holes  exchanged.

The residue $Z$ of the poles goes to zero as
\begin{equation}
Z \sim (k_\perp)^{ {1 \over 2 \nu_{k_F}} -1 } \approx k_\perp^{1.089} \,,
\end{equation}
and the dispersion relation goes like
\begin{equation}
\omega_* \sim (k_\perp)^{ {1 \over 2 \nu_{k_F}} } \approx k_\perp^{2.089} \,.
\end{equation}

\section{Fermi surfaces from five-dimensional supergravity}

We turn now to the case of five-dimensional ${\cal N}=8$ gauged supergravity \cite{Gunaydin:1985cu}, where the gauge group is $SO(6)$, and the simplest choice of gauge fields that allows the scalars to remain in their trivial configuration (corresponding to an undeformed $S^5$ in a lift to type IIB supergravity) is
\begin{equation}\label{5DGaugeField}
a_\mu \equiv A_{\mu12} = A_{\mu34}=  A_{\mu56}\,. 
\end{equation}
A consistent truncation of the bosonic lagrangian, omitting a Chern-Simons term which will not figure in our discussion, is
\begin{equation}
e^{-1} {\cal L} = -{1 \over 4} R - {3 \over 4} f_{\mu\nu} f^{\mu\nu} + {3  g^2 \over 4} \,.
\end{equation}
The RNAdS solution is again given by (\ref{RNAdS}), (\ref{HorizonFunction}) with $d=4$ and
\begin{equation}
g = {2 \over L} \,, \qquad\qquad \mu = {r_0 \over \sqrt{2} L} \,,
\end{equation}
matching the notation of \cite{Faulkner:2009wj} with $g_F = L/\sqrt{3}$.

We turn now to the Fermi fields.  Their index structure is based on a local symmetry group $USp(8)$ whose fundamental $8$-dimensional representation splits into the ${\bf 4} + \overline{\bf 4}$ of $SO(6)$.  There are 48 spin-1/2 fermions $\chi_{abc} = \chi_{[abc]|}$ where $a,b,c = 1 \ldots 8$ are $USp(8)$ indices and $[\ldots]|$ denotes antisymmetrization with the symplectic trace removed.  The $\chi_{abc}$ obey the symplectic Majorana condition, a condition particular to five dimensions relating one field to the complex conjugate of another via the $USp(8)$ symplectic metric $\Omega_{ab}$.  A somewhat involved calculation starting from the lagrangian quoted in \cite{Gunaydin:1985cu} leads to the following quadratic action for the spin-$1/2$ fields:
\begin{eqnarray}\nonumber
&& e^{-1} {\cal L}_{1/2} = {i \over 12} \bar\chi^{abc} \gamma^\mu \nabla_\mu \chi_{abc} \\ && + {ig  \over 16} \bar\chi^{abc} \delta_{cd}   \Omega^{de} \chi_{abe}  - {i g \over 8 } \bar\chi^{abc} \gamma^\mu a_\mu \tilde\Gamma^{cd} \chi_{abd}\\
&& + {i \over 16 } f_{\mu\nu} \tilde\Gamma^{ab}  \left( \sqrt{2} \bar\psi^c_\rho \gamma^{\mu\nu} \gamma^\rho \chi_{abc} + \bar\chi_a^{\;\;cd} \gamma^{\mu\nu} \chi_{bcd}  \right) \,.
\nonumber\end{eqnarray}
Here $\tilde\Gamma \equiv \Gamma_{12} + \Gamma_{34} + \Gamma_{56}$, and the $\Gamma_{IJ}\equiv [\Gamma_I, \Gamma_J]/2$ are elements of the $SO(6)$ Clifford algebra.  These gamma matrices are essentially Clebsch-Gordan coefficients relating $SO(6)$ and $USp(8)$.

Under the inclusion $USp(8) \supset SO(6)$, the 48 fermions organize themselves into complex symplectic Majorana pairs in the ${\bf 20} + {\bf 4}$, while the gravitini are in the ${\bf 4}$. 
In terms of the $U(1) \subset SO(6)$ defined by the gauge field $a_\mu$ (\ref{5DGaugeField}),  one can analyze the weight vectors to see that the ${\bf 20}$ contains 3 elements with charge $5/2$, 8 elements with charge $3/2$ and 9 elements with charge $1/2$, while the ${\bf 4}$ contains an element of charge $3/2$ and three of charge $1/2$.  (Here and below we indicate the charge magnitude, since each symplectic Majorana pair contains excitations of both signs of the charge.) 
  Thus the three complex fermions with charge $5/2$ cannot mix with the gravitini; it is these that we will study.

We were able to find simultaneous eigenvectors of the kinetic, mass, gauge and Pauli operators corresponding to these three complex fermions; each obeys the same decoupled Dirac equation,
\begin{equation}
\left( i\gamma^\mu \nabla_\mu \pm {5  \over L}\gamma^\mu a_\mu  \mp {1 \over 2L}  + {i \over 4} f_{\mu\nu} \gamma^{\mu\nu} \right) \chi = 0 \,,
\end{equation}
where the lower set of signs is for the conjugate excitation in the symplectic Majorana pair.
In solving the conjugate equation one may for convenience switch the sign of the gamma matrices, effectively flipping the signs of both $m$ and the Pauli term; the conjugate then has the same mass but opposite signs in both couplings to the gauge field, and is thus equivalent to studying the original fluctuation in the the background of an oppositely-charged black hole.

Considering solutions to this equation for either sign charge, the near-boundary scaling is controlled by the mass $mL = 1/2$ and takes the form,
\begin{eqnarray}
\chi_+ &\sim& A r^{-3/2} + B r^{-7/2} \,, \\
\chi_- &\sim& C r^{-5/2} \log r+ D r^{-5/2} \,,
\nonumber
\end{eqnarray}
for spinors $\chi_\pm$ that are eigenvectors of $\gamma^r$, where again $A$ is the source and $D$ the response.  

Pauli couplings were considered in \cite{Edalati:2010ge}, and generalize the formulas of  \cite{Faulkner:2009wj} for the near-horizon behavior of the solutions by shifting the momentum $k \to \tilde{k}$, where in our case
\begin{equation}\label{ShiftedMomentum}
\tilde{k} \equiv k \mp {r_0 \over \sqrt{2} L^2} \,,
\end{equation}
where the different signs correspond to different eigenvectors of $\gamma^r \gamma^t \vec{k} \cdot \vec{\gamma}$; each of $\chi_+$ and $\chi_-$, with two components each, contains one component with each sign shift.

This leads
to near horizon behavior of the form (\ref{Horizon}) with
\begin{eqnarray}\label{nukFive}
\nu_k 
= {\sqrt{47} \over 12} \sqrt{{150 \over 47}\left( \tilde{k} \over  \mu q\right)^2 - 1} \,. 
\end{eqnarray}
The oscillatory region of imaginary $\nu_k$ thus occurs for $|\tilde{k}| < 
\sqrt{47/150} (\mu q)$.

Selecting infalling boundary conditions and again taking $L = r_0 = 1$ for simplicity, we find Fermi surfaces with
\begin{equation}
\tilde{k}_F \approx \pm 2.00000 \,,  \label{kFTwo}
\end{equation}
or $\tilde{k}_F/(\mu q) \approx \pm 0.56569$.
(Note the values of $k_F$ will have opposite signs for different components due to (\ref{ShiftedMomentum}).)
The precision of this result is suggestive that the $\tilde{k}_F$ values are $\pm 2$ exactly; notice that the Fermi surfaces are barely outside the oscillatory region.  The corresponding value of $\nu_k$ is then exactly
\begin{equation}\label{5DNu}
\nu_{k_F} = {1 \over 12} \,.
\end{equation}
Thus these Fermi surfaces also have $\nu_{k_F} < 1/2$, meaning again we have a non-Fermi liquid.  

Again the retarded Green's function takes the form (\ref{GreensFunction}).  
For negative $q$ we have $\gamma_F  \approx 0.0126$, whose small size makes $e^{i\gamma_F}$  nearly a real number, and consequently again 
the fermionic quasinormal frequency $\omega_{\rm QNM} = \omega_* - i\Gamma$ lying closest to the Fermi surface is nearly real:
\begin{equation}
 {\Gamma \over \omega_*} = \tan 6 \gamma_F \approx {1 \over 13.2}\,, \label{SmallWidth}
\end{equation}
for both $k_\perp > 0$ and $k_\perp < 0$.  Again there are no stable quasiparticles, but some modes have a small ratio of width to excitation energy.  Unlike the four-dimensional case, we find a symmetry between particles and holes, which can be traced to the rational value of $\nu_{k_F}$ (\ref{5DNu}).  Positive $q$ again sees the same physics with particles and holes exchanged.

Moreover we find an unusual sixth-order dispersion relation,
\begin{equation}
 \omega_* \propto (k_\perp)^{1 \over 2 \nu_{k_F}} = k_\perp^6 \,,  \label{SixthOrder}
\end{equation}
where the constant of proportionality involves $h_2$, as well as a rapidly vanishing residue,
\begin{equation}
Z \sim (k_\perp)^{{1 \over 2 \nu_{k_F}}-1} =  k_\perp^5 \,.
\end{equation}

\section{Field theory operators and charge density}

 In five dimensions, where the field theory dual is ${\cal N}=4$ super-Yang-Mills theory, the operators dual to the three charge $5/2$ fermion fields are ${\cal O}_j = \tr \lambda Z_j$ for $j = 1,2,3$, where $\lambda$ is the charge $3/2$ gaugino in the ${\bf 4}$ of the $R$-symmetry group $SO(6)$, and $Z_j = X_{2j-1}+iX_{2j}$ where $X_I$ are the six adjoint scalars.  Thus the gaugino contributes $+3/2$ to the total charge of ${\cal O}_j$, while the scalars contribute $+1$.

A common view 
in the current literature is that the singularity in the two-point function of ${\cal O}_j$ at $\omega=0$ and $k=k_F$ is a signal of a Fermi surface formed by color singlet bound states of $\lambda$ and $Z_j$, which we will term mesinos: see for example \cite{Huijse:2011hp,Iqbal:2011ae,Huijse:2011ef}. 
The two-point function we find here scales as
$N^2$,  suggestive of a sum over all possible gaugino colors and hence an alternative interpretation \cite{Gubser:2009qt}, that this singularity 
is due to a Fermi surface formed by the adjoint gauginos $\lambda$ themselves, perhaps
dressed in some way by  strong gauge interactions.

A Luttinger count of the charge density due to the charged fermions is
 \begin{eqnarray}
  j_{\rm fermions} &=& \sum_{\rm Fermi\ surfaces} q_{\rm f} g_{\rm s} \int_{|k| < |k_F|} 
     {d^3 k \over (2\pi)^3}  \nonumber \\
     &=& {q_{\rm f} \over 6\pi^2} \left( g_+ |k_{F,+}|^3 + g_- |k_{F,-}|^3 \right) \,, 
    \label{jFermion} 
 \end{eqnarray}
where $q_{\rm f}$ is the dimensionless charge of the gauge theory fermion involved and $g_{\rm s}$ is a degeneracy factor indicating the number of distinct fermions participating in the Fermi surface dynamics.  In the second line of (\ref{jFermion}) we have specialized to the case of two Fermi surfaces, the number found in our analysis, with degeneracy factors $g_+$ and $g_-$.  In the gaugino interpretation, clearly $q_{\rm f} = 3/2$, and with $L=r_0=1$, the result (\ref{kFTwo}) indicates that $|k_{F,\pm}| = 2 \pm 1/\sqrt{2}$.  For the degeneracy factors, the most natural conjecture is that $g_+ = g_- = N^2$, which is simply the number of colors at leading order in $N$.  There is no additional factor of $2$ for the spin of the fermions because the gauginos are chiral.  
We then arrive at
 \begin{equation}
  j_{\rm fermions} = {11 \over 2\pi^2} N^2 \,.  \label{FoundjFermion}
 \end{equation}  
This is 
numerically quite close to the total charge of the black brane,  
 which referring to \cite{Cvetic:1999rb} we calculate to be
 \begin{equation}
  j_{\rm total} = {9 \sqrt{6} \over 4\pi^2} N^2 \,, \label{jTotal}
 \end{equation}
with $L=r_0=1$ as before;
this is related to the scaled quantities $j_i$ of \cite{Cvetic:1999rb} (where our $r_0$ is denoted $r_H$) by $j_{\rm total} = N^2 (j_1 + j_2 + j_3)$.  Comparing (\ref{FoundjFermion}) and (\ref{jTotal}), we see that
 \begin{equation}
  {j_{\rm fermions} \over j_{\rm total}} = \sqrt{242 \over 243} \,. \label{jRatio}
 \end{equation}
Taking (\ref{jRatio}) at face value, we should conclude that most of the charge is indeed carried by the charge $+3/2$ gauginos.  Let us bear in mind, however, that the choice $g_+ = g_- = N^2$ was conjectural.  More conservatively, we could regard (\ref{jRatio}) as an upper limit on the amount of charge carried by these gauginos.  
See \cite{Hartnoll:2010xj,Huijse:2011hp,Huijse:2011ef} for somewhat different approaches to the Luttinger theorem in holographic (non-) Fermi liquids.

There are other positively charged particles in the field theory, namely charge $+1/2$ gauginos and charge $+1$ scalars.  If (\ref{jRatio}) is correct, one should ask why they carry such a small fraction of the charge.   The charge $+1/2$ gauginos may not carry appreciable charge because there are tree-level interactions that convert pairs of them into charge $+1$ bosons, whereas no such interactions exist to convert two charge $+3/2$ gauginos into charge $+3$ bosons, simply because there are no charge $+3$ bosons.\footnote{We thank D.~Huse for suggesting this line of reasoning to us.} 

Charge  carried by scalars, if not in a mesino Fermi surface, is likely
in the form of a Bose condensate.  Assuming the scalars do condense, they must do so in a manner that preserves the $U(1)$ gauge symmetry, since this symmetry is obviously unbroken in the RNAdS${}_5$ solution.  This is possible in the large $N$ limit, since the $N$ D3-branes can be distributed continuously over a $U(1)$-invariant configuration---perhaps even an $SO(6)$-invariant configuration.   In fact, highly symmetric, continuous distributions of D3-branes have been previously derived as limits of spinning brane solutions \cite{Kraus:1998hv,Freedman:1999gk}. 
If the scalars condense in such a fashion, 
the operator ${\cal O}_j = \tr \lambda Z_j$ may have a finite amplitude to create a zero momentum boson $Z_j$ and to put all its momentum into the charge $+3/2$ gaugino $\lambda$.  This is just what we need in order to explain why a Fermi surface for the gauginos would give rise to the singularity we observe in the ${\cal O}(N^2)$ two-point function of ${\cal O}_j$. 
Thus, a gaugino Fermi surface together with a symmetry-preserving scalar condensate appears to be a plausible alternative to the mesino interpretation of holographic Fermi surface singularities.  However, we cannot offer any line of reasoning to explain why the fraction of the charge carried by the scalars should be as small as (\ref{jRatio}) suggests.  The ${\cal O}(N^2)$ zero-point entropy of the RNAdS${}_5$ solution remains a puzzle, and an understanding of the entanglement entropy in the presence of holographic Fermi surfaces is as yet incomplete; see however \cite{Ogawa:2011bz} for work in this direction.

In the $AdS_4$ case, the field theory dual is more complicated than ${\cal N}=4$ super-Yang-Mills \cite{Schwarz:2004yj,Bagger:2007jr,Gustavsson:2007vu,Aharony:2008ug}, and the overall scaling of Green's functions is $N^{3/2}$.  In the theory of a single M2-brane, the operators dual to the charged fermions again have the form $X\lambda$.  In the standard field theory presentation of $SO(8)$, $X$ is a vector and $\lambda$ is a spinor.
Under the $U(1)$ which is the diagonal sum of all four Cartan generators, the scalars and the fermions both have charge $\pm 1$.  With a little care one can see that $15$ combinations $X\lambda$ have charge $+2$, another $15$ have charge $-2$, and the rest are neutral.  The $15+15$ charged combinations exactly match the $30$ charged components of $\chi_{ijk}$.  While this indicates that all the fundamental fermions in the M2-brane theory are involved equally in the Fermi surface revealed by our $AdS_4$ gauged supergravity calculations, we have not formulated a Luttinger count analogous to (\ref{jFermion}) and (\ref{FoundjFermion}).

\section*{Acknowledgments}

We are grateful to D.~Huse and S.~Kachru for helpful discussions, A.~Yarom for comments on the manuscript, and S.~Sachdev for helping motivate the work.  The work of O.D.\ and C.R.\ was supported by the Department of Energy under Grant No.~DE-FG02-91-ER-40672.  The work of S.S.G.~is supported in part by the Department of Energy under Grant No.~DE-FG02-91ER40671.

\bibliography{maximal}
\end{document}